\documentclass{kapproc}
\usepackage{graphicx}
\newcommand{\bm}[1]{{\mathbf{#1}}}
\newcommand{\bbox}[1]{{\mathbf{#1}}}
\newcommand{\onlinecite}[1]{\cite{#1}}
\normallatexbib

\begin{document}

\articletitle[Fast control in quantum dots]{Fast control 
of quantum states in quantum dots: \\limits due to decoherence}
\author{Lucjan Jacak, Pawe{\l} Machnikowski}
\affil{Institute of Physics, Wroc{\l}aw University of Technology,
50-370 Wroc{\l}aw, Poland}
\author{Jurij Krasnyj}
\affil{Institute of Mathematics, University of Opole, 45--051 Opole, 
Poland\thanks{On leave from Odessa University, Ukraine}}

\begin{abstract}
We study the kinetics of confined carrier-phonon system in a quantum dot
under fast optical driving and discuss the resulting limitations to fast
coherent control over the quantum state in such systems.
\end{abstract}

\section{Introduction}

Unlike natural atoms, semiconductor quantum dots (QDs) always form
part of a macroscopic crystal. The interaction with the quasi-continuum of
lattice degrees of freedom (phonons) constitutes an inherent feature 
of these nanometer-size systems and cannot be neglected in any realistic 
modeling of QD properties, especially when the coherence of confined 
carriers is of importance. The understanding of the decisive role played 
by the QD coupling to lattice modes has increased recently due to both
experimental and theoretical study (spectrum reconstruction
\cite{hameau99,hameau02,verzelen02a,jacak03a}, 
relaxation \cite{heitz97,heitz01,ignatiev01,verzelen00,jacak02a,verzelen02b},
phonon replicas and phonon-assisted transitions 
\cite{heitz99,heitz00,lemaitre01,findeis00,fomin00,jacak02f}, 
phonon-induced pure dephasing	upon ultrafast excitation 
\cite{borri01,borri02a,krummheuer02,vagov02a,vagov03,jacak03b}).
The phonon-induced
decoherence seems to be crucial for any quantum information processing 
application and for any nanotechnological device relying on quantum 
coherence of confined carriers \cite{alicki03,machnikowski03a}.

There are three major mechanisms of carrier-phonon interaction 
\cite{mahan00}: (1) Coulomb
interaction with the lattice polarization induced by the relative
shift of the positive and negative sub-lattices of the polar compound,
described upon quantization by longitudinal optical (LO) phonons; 
(2) deformation potential coupling describing the band shifts due to 
lattice compression, i.e. longitudinal acoustical (LA) phonons;
(3) Coulomb interaction with piezoelectric field generated by shear crystal
deformation (transversal acoustical, TA, phonons). The lattermost effect
is weak in InAs/GaAs systems but may be of more importance for the
properties e.g. of GaN dots 
\cite{hohenester99c,hohenester00d,derinaldis02}.

Due to the carrier-phonon interaction, any change in the carrier
subsystem must be accompanied by the appropriate modification of the
lattice state. An example of this effect may be creation of the
exciton-polaron states: excitons accompanied by the lattice polarization
(coherent LO phonon field) with energy
shifted down by a few meV \cite{verzelen02a,jacak03b} . The
interaction with acoustical branches leads to creation of a similar
deformation field which, although much less pronounced in terms of
energy shift, strongly influence the dynamical properties of the
interacting system due to gapless spectrum and low characteristic frequencies
of the acoustical phonons.

The lattice response to any manipulation performed on the charge
distribution confined in a QD leads to discrepancy between the desired
state of the quantum confined system and the actual one. 
This effect may be stronger than any decoherence process in an undriven
system. In terms of the
quantum information processing schemes, such a discrepancy is manifested
by the loss of fidelity of the quantum operation. In quantum
nano-technology applications, it will reduce the efficiency of
nano-devices.

In the present paper we discuss the carrier-lattice kinetics induced by
optical excitation of a confined exciton. 
The paper is organized as follows: In the next section we present the
formal model of the exciton confined in a QD.
The section \ref{sec:dress} describes the
system kinetics in response to an abrupt change of the charge
distribution. In the section \ref{sec:trade} we discuss the trade-off
between the dynamically induced error and other processes limiting the
coherence time of a quantum state. The final section contains concluding
discussion.

\section{The model}

We consider the Hamiltonian describing a single exciton interacting with
phonons,
\begin{eqnarray*}
H &  = &  \sum_{n}\epsilon_{n}a^{\dagger}_{n}a_{n}
+\sum_{{\bbox{k}},s} \omega_{s,{\bbox k}}
b_{s,{\bbox k}}^{\dagger}b_{s,{\bbox k}} \\
& & +\frac{1}{\sqrt{N}}\sum_{{\bbox{k}},n,n',s}
F^{(s)}_{n,n'}( {\bbox k}) a_{n}^{\dagger}a_{n'}
\left( b_{s,{\bbox k}}+b_{s,{-\bbox k}}^{\dagger} \right),
\end{eqnarray*}
where 
$b_{s,{\bbox k}},b^{\dagger}_{s,{\bbox k}}$ are bosonic annihilation
and creation operators for LO ($s=\mathrm{o}$) and LA ($s=\mathrm{a}$)
phonons with quasi-momentum ${\bbox k}$. The corresponding frequencies
within the effectively coupled wavevector range may be modeled by
$\omega_{\mathrm{o},{\bbox k}}=\Omega_{\bbox k}\simeq \Omega -\beta
k^{2}$ and  $\omega_{\mathrm{a},{\bbox k}}=ck$.  
The coupling constants for the LO and LA phonon branches are given
by
\begin{displaymath}
F^{(\mathrm{o})}_{n,n'}( \bbox k)=-
\frac{e}{k}
\sqrt{\frac{2\pi\Omega}{v\tilde{\epsilon}}}
\int\Phi^{*}_{n} (\bbox{r}_{\mathrm{e}}, \bbox{r}_{\mathrm{h}})
\left(e^{i\bbox{k}\cdot\bbox{r}_{\mathrm{e}}}
- e^{i\bbox{k}\cdot\bbox{r}_{\mathrm{h}}}\right)
\Phi_{n'}(\bbox{r}_{\mathrm{e}},
\bbox{r}_{\mathrm{h}})d^{3}{\bbox{r}_{\mathrm{e}}}
d^{3}{\bbox{r}_{\mathrm{h}}}
\end{displaymath}
and
\begin{eqnarray}
\lefteqn{F^{(\mathrm{a})}_{n,n'}( {\bbox k})=} \label{fa} \\
& & -\sqrt{\frac{k}{2\varrho v c}}
\int\Phi^{*}_{n} (\bbox{r}_{\mathrm{e}}, \bbox{r}_{\mathrm{h}})
\left(\sigma_{\mathrm{e}}e^{i\bbox{k}\cdot\bbox{r}_{\mathrm{e}}}  -
\sigma_{\mathrm{h}}e^{i\bbox{k}\cdot\bbox{r}_{\mathrm{h}}}\right)
\Phi_{n'}(\bbox{r}_{\mathrm{e}}, \bbox{r}_{\mathrm{h}})
d^{3}\bbox{r}_{\mathrm{e}} d^{3}\bbox{r}_{\mathrm{h}},
\nonumber
\end{eqnarray}
where $\bbox{r}_{\mathrm{e}}, \bbox{r}_{\mathrm{h}}$ denote
the coordinates of the electron and hole, respectively, 
and $\Phi_{n}(\bbox{r}_{\mathrm{e}}, \bbox{r}_{\mathrm{h}})$ is the exciton 
wavefunction.
The other elements of the notation are described in 
Table \ref{tab:param}, along with
values (corresponding to InAs/GaAs system) used in the calculations.

\begin{table}[tb]
\begin{tabular}{lll}
\hline
Electron mass & $m_{\mathrm{e}}$	& $0.067m_{0}$ \\
Hole mass & $m_{\mathrm{h}}$	& $0.38m_{0}$ \\
Static dielectric constant & $\varepsilon_{\mathrm{s}}$ & 13.2 \\ 
Effective dielectric constant & $\tilde{\epsilon}$ & 62.6 \\ 
Optical phonon energy & $\Omega_{0}$ & $36$ meV \\
Longitudinal sound speed & $c$ & 5150 m/s \\
Deformation potential for electrons & $\sigma_{\mathrm{e}}$ & $6.7$ eV \\
Deformation potential for holes & $\sigma_{\mathrm{h}}$ & $-2.7$ eV \\
Unit crystal cell volume & $v$ & 0.044 nm$^{3}$ \\
LO phonon dispersion parameter & $\beta$ & 0.03 meV$\cdot$nm$^{2}$\\
\hline
\end{tabular}
\caption{\label{tab:param}The material parameters used in the
calculations (partly after Refs. \protect\onlinecite{adachi85,strauch90}).}
\end{table}

\begin{figure}[tb]
\unitlength 1mm
\begin{center}
\begin{picture}(85,58)
\put(0,0){\resizebox{85mm}{!}{\includegraphics{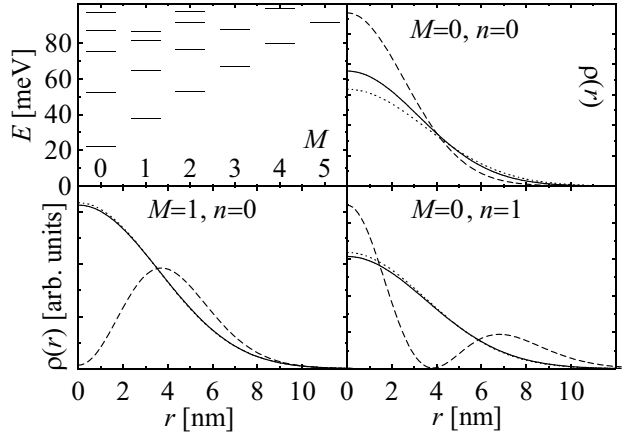}}}
\end{picture}
\end{center}
\caption{\label{fig:x-spectr}The spectrum of the exciton in QD (upper
left) and the electron (solid lines) and hole (dashed
lines) probability densities for the three lowest exciton states,
compared to the ground state
probability density for a noninteracting particle (dotted lines). Here
the noninteracting particle wavefunction width is $l_{\perp}=4.9$ nm
in-plane and $l_{z}=2$ nm in the growth direction.}
\end{figure}

Numerical diagonalization of the interacting electron--hole system in 
parabolic confinement, under assumption that non-interacting
electron and hole would have the same wavefunctions,
leads to the spectrum shown in Fig. \ref{fig:x-spectr}a ($M$ is the
conserved total angular momentum).
The dominant contribution to the
lowest excited states of the exciton comes from the excited hole states,
while the electron wavefunction is only slightly modified.
The corresponding electron and hole distributions are shown in 
Fig. \ref{fig:x-spectr}b-d. 

\begin{figure}[tb]
\unitlength 1mm
\begin{center}
\begin{picture}(85,58)
\put(0,0){\resizebox{85mm}{!}{\includegraphics{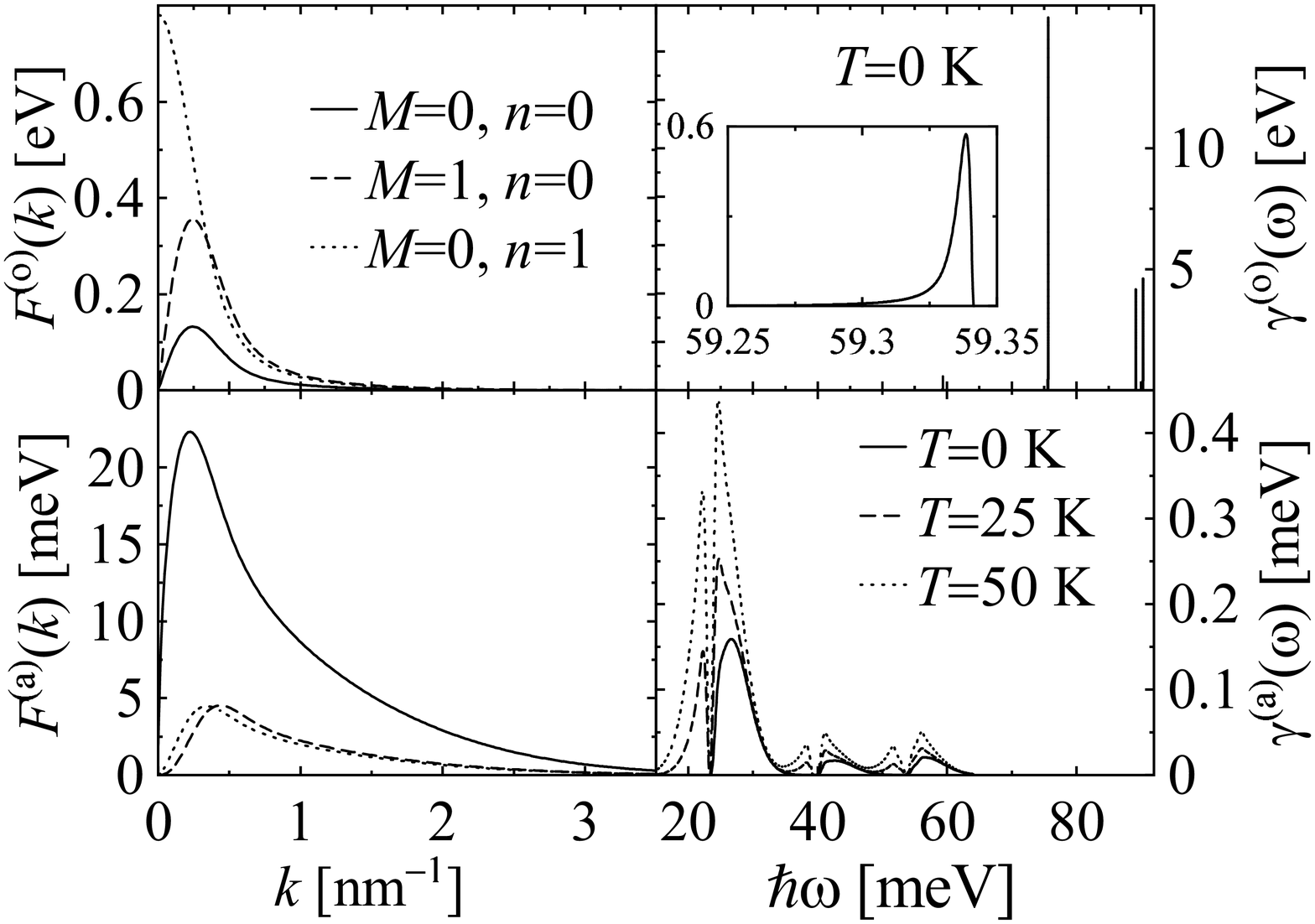}}}
\end{picture}
\end{center}
\caption{\label{fig:fnn} Coupling constants for LA (a) and LO (b) phonon
modes between the ground state and the three lowest states and the
resulting structure of the imaginary part of the mass operator (c,d).}
\end{figure}

The exciton-phonon coupling functions may be calculated using the 
wavefunctions found numerically. Fig.
\ref{fig:fnn} (left panels) shows the results for coupling between 
the ground state and a few lowest states, averaged over angles.
The coupling to LO phonons is much stronger than to LA phonons. 
In the case of LO phonons, the coupling strength increases for
excited states due to less
charge cancellation between the electron and hole in these states (cf.
Fig. \ref{fig:x-spectr}b-d).

\section{System kinetics after an ultrafast pulse}
\label{sec:dress}

Let us study the kinetics of the system after an excitation performed by
an extremely short (formally infinitely short) laser pulse.
The information on the dynamics of the interacting carrier--lattice
system is contained in the exciton single-particle causal Green function,
\begin{displaymath}
G_{n,n'}(t)= -i
\langle T\{a_{n}(t) a^{\dagger}_{n'}(0)\}\rangle.
\end{displaymath}
The  averaging $\langle \ldots\rangle $ is the temperature-dependent 
averaging with respect to
phonon degrees of freedom and the vacuum of exciton (cf. Ref.
\onlinecite{suna64}; it corresponds to the case when
the grand canonical averaging sector without exciton, vacuum, is
energetically distant, $\sim 1$ eV, from the next sectors).
For $t\geq 0$ this Green function coincides (up to a constant factor)
with the correlation function
$\langle a_{n}(t) a^{\dagger}_{n'}(0)\rangle $ which, for $n_{1}=n_{2}$
may be interpreted as a measure of integrity of the excitonic state:
it corresponds to the overlap of the state at time $t$ 
with this state at the initial moment $t=0$.
The Fourier transform of the correlation function,  $I_{n,n'}(\omega)=
\int_{-\infty}^{\infty} \langle a_{n}(t)a^{\dagger}_{n'}(0)\rangle e^{i\omega t}dt$,
is usually called the spectral density \cite{abrikosov75,bonch62},
and it can be expressed by the imaginary part of the causal Green function:
\begin{equation}
{\mathrm{Im}} G_{n,n'}(\omega)
=-\frac{1}{2}\left(1+e^{-\omega/k_{B}T}\right)^{-1}
I_{n,n'}(\omega),
\label{ImGc-cor}
\end{equation}
where
$G_{n,n'}(\omega)
=\int_{-\infty}^{\infty}G_{n,n'}(t)e^{i\omega t}dt$.

The equation of motion for the causal Green function 
may be rewritten as a Dyson--type equation, 
\begin{equation}
( \omega - \epsilon_{n}) G_{n,n'}(\omega)
- \sum _{n_{3}}
M_{n, n''} (\omega)   G_{n'',n'}(\omega)
=\delta_{n,n'},
\label{2}
\end{equation}
with the mass operator
\begin{eqnarray*}
M_{n,n'}(\omega) & = & \frac{i}{2\pi N}
\sum_{\bbox k, s,n'',n'''}  F^{(s)}_{n,n''}(\bbox k)
F^{(s)}_{n''',n'}(-\bbox k) \\
& &
\times\int d\omega' G_{n'',n'''}(\omega+ \omega')
D^{0}(\bbox k, s ,\omega'),
\end{eqnarray*}
where we have restricted ourselves to single-phonon processes by replacing
the vertex function by its zeroth order approximation and using the free
phonon Green function (LO phonon damping may be easily included later: see
below).
Since $F^{(s)}_{n,n'}(\bbox k) \sim g_{s}$, then
$|M_{n,n'}(\omega)|^{2} \sim g_{s}^{2}$. Hence, up to 
$g_{s}^{2}$ one has from Eq. (\ref{2}),
\begin{equation}
G_{n,n}(\omega)=
\frac{1}{\omega -\epsilon_{n} -M_{n,n}(\omega)}
\label{gc}
\end{equation}
and 
\begin{eqnarray}
M_{nn}(\omega) & = & \Delta_{n}(\omega)-i\gamma(\omega) \label{massop} \\
& = & \frac{i}{2\pi N}\sum_{n',\bbox k} |F_{nn'}^{(s)}({\bbox k})|^{2}
\int d\omega' G_{n'n'}(\omega+\omega')D^{(0)}(s,{\bbox k},\omega').
\nonumber
\end{eqnarray}

For the real and the imaginary part of the
mass operator we obtain form (\ref{massop})
\begin{eqnarray}
\Delta_{n}(\omega)& = & 
\frac{1}{N}
\sum_{\bbox k,s,n,}
| F^{(s)}_{nn'}(\bbox k)|^{2}  \label{delta-big} \\
& & \times\left[
\frac{(1+n_{s,{\bbox k}})
(\omega -\epsilon_{n'} -\Delta_{n'}(\omega-\omega_{s,{\bbox k}})
-\omega_{s,{\bbox k}})}
{[\omega -\epsilon_{n'} -\Delta_{n'}(\omega-\omega_{s,{\bbox k}})-
\omega_{s,{\bbox k}}]^{2}+\gamma^{2}_{n'}(\omega-\omega_{s}
({\bbox k}))}\right. \nonumber \\
& & \left.
+\frac{n_{s,{\bbox k}}
(\omega -\epsilon_{n'} -\Delta_{n'}(\omega+\omega_{s,{\bbox k}})
+\omega_{s,{\bbox k}})}
{[\omega -\epsilon_{n'} -\Delta_{n'}(\omega+\omega_{s,{\bbox k}})+
\omega_{s,{\bbox k}}]^{2}
+\gamma^{2}_{n'}(\omega+\omega_{s,{\bbox k}})}\right] \nonumber
\end{eqnarray}
and
\begin{eqnarray}
\gamma_{n}(\omega) & = & \frac{1}{N}
\sum_{\bbox k,s,n'}
| F^{(s)}_{nn'}(\bbox k)|^{2} \label{31b} \\
& & \times \left[
\frac{(1+n_{s,{\bbox k}})
\gamma_{n'}(\omega-\omega_{s,{\bbox k}})}
{[\omega -\epsilon_{n'} -\Delta_{n'}(\omega-\omega_{s,{\bbox k}})-
\omega_{s,{\bbox k}}]^{2}+\gamma^{2}_{n'}(\omega-\omega_{s}
({\bbox k}))}\right.
\nonumber \\
& & \left.
+\frac{n_{ s,{\bbox k}}\gamma_{n'}(\omega+\omega_{s,{\bbox k}})
}
{[\omega -\epsilon_{n'} -\Delta_{n'}(\omega+\omega_{s,{\bbox k}})+
\omega_{s,{\bbox k}}]^{2}
+\gamma^{2}_{n'}(\omega+\omega_{s,{\bbox k}})}\right]. \nonumber
\end{eqnarray}

In this equation the first term  determines the energy
shift due to dressing with LO phonons, while the second one corresponds to
LA phonons. 
The former dominates energetically: the energy shift
is mainly due to the interaction of exciton with optical phonons 
(dressing of exciton with LO phonons, creating together an exciton-polaron).
The latter is
small and its energetical effect may be safely neglected but it 
contributes considerably to the system kinetics.

To the lowest order, one may neglect $\Delta_{n'}$ and $\gamma_{n'}$
in the right-hand-side of (\ref{delta-big}) and, neglecting the acoustic
phonon term, write the equation for the exciton-polaron energy levels
$E_{n}$ as the poles of the Green function (\ref{gc}), 
$ E_{n}-\epsilon_{n}-\Delta_{n}(E_{n})=0$, i.e.
\begin{equation}
E_{n} - \epsilon_{n}
-\frac{1}{N} \sum_{\bbox k,n}
|F^{(\mathrm{o})}_{nn}(\bbox k)|^{2}
\left[\frac{1+n_{\mathrm{o},{\bbox k}}}{E_{n} -\epsilon_{n} -\Omega}
+\frac{n_{\mathrm{o},{\bbox k}}}{E_{n} -\epsilon_{n} +\Omega}\right] =0.
\label{del}
\end{equation}
At $k_{\mathrm{B}}T\ll\Omega$, 
the above equation is equivalent to that found by Davydov 
diagonalization of the Fr\"ohlich Hamiltonian for exciton \cite{davydov72}. 
The Eq. (\ref{del}) for $n=0$ gives the ground
state energy shift $\Delta_{0}\sim -5 $ meV for the QD with parameters 
listed in Table \ref{tab:param} (see Ref. \onlinecite{jacak03b}).

The imaginary part of the mass operator is given to the leading order 
by the equation (putting $\gamma_{n'} =0$ in the rhs. of Eq. (\ref{31b})):
\begin{eqnarray}
\lefteqn{\gamma_{n}(\omega) =} \label{gamn} \\ & & 
\!\!\!\!\frac{\pi}{N}\sum_{\bbox k, n}\left\{
|F^{(\mathrm{o})}_{nn'}(\bbox k)|^{2}
\left[(1+n_{\mathrm{o},{\bbox k}})
\delta(\omega -E_{n'} -\Omega_{\bbox k})
+n_{\mathrm{o},{\bbox k}}
\delta(\omega -E_{n'} +\Omega_{\bbox k})\right]\right. \nonumber \\
& & \left. +|F^{(\mathrm{a})}_{nn'}(\bbox k)|^{2}
\left[(1+n_{\mathrm{a},{\bbox k}})\delta(\omega
-E_{n'} - c k)
+n_{\mathrm{a},{\bbox k}}\delta(\omega -E_{n'}
+ ck)\right]\right\}, \nonumber
\end{eqnarray}
where we use the fact that the equation 
$\omega-\epsilon_{n'}-\Delta_{n'}(\omega\pm\omega_{s, {\mathbf k}})
\pm\omega_{s, {\mathbf k}}=0 $
is solved by $\omega=E_{n'}\mp \omega_{s, {\mathbf k}}$ and neglected 
higher-order corrections resulting from resolving the Dirac $\delta$.
The first term in Eq. (\ref{gamn}) describes the energy transfer to the
LO phonon sea, while the
second one corresponds to the energy transfer
form gradually dressing exciton  to the LA phonon sea.

The form of $\gamma_{0}(\omega)$ for our model, including 
a few lowest states, obtained using the numerical wavefunctions
is shown for various temperatures in Fig \ref{fig:fnn}c,d.

Using (\ref{gc}), one finds 
\begin{equation}
{\mathrm{Im}} G_{n,n}(\omega)=
\frac{\gamma_{n}(\omega)}{[\omega-\epsilon_{n}-\Delta_{n}(\omega)]^{2}
+\gamma_{n}^{2}(\omega)}.
\label{ImG}
\end{equation}
In the following, we will focus on the ground exciton state, $n=0$.
Usually, this state is long-living at low temperatures: the broadening of
the corresponding spectral line, related to radiative lifetime and
thermally activated processes, does not exceed 0.1 meV for $T<100$ K
\cite{bayer02,borri01}. 

Let us note that the term $\gamma_{0}^{2}(\omega)$ in the denominator
of (\ref{ImG}) is
important only near $E_{0}$, where the other term vanishes (otherwise, it
is a higher-order correction). 
Therefore, it may be approximated by a
constant $\gamma_{0}=\gamma_{0}(E_{0})$.
In a similar manner, the correction $\Delta_{0}(\omega)$ may be replaced
by 
$\Delta_{0}(\omega)\approx \Delta_{0}(E_{0})
+(\omega-E_{0}) (d\Delta_{0}(\omega)/d\omega)_{\omega=E_{0}}$
and combined with $\epsilon_{0}$ to give
$E_{0}$, in accordance with (\ref{del}). 

\begin{figure}[tb]
\unitlength 1mm
\begin{center}
\begin{picture}(85,60)
\put(0,0){\resizebox{85mm}{!}{\includegraphics{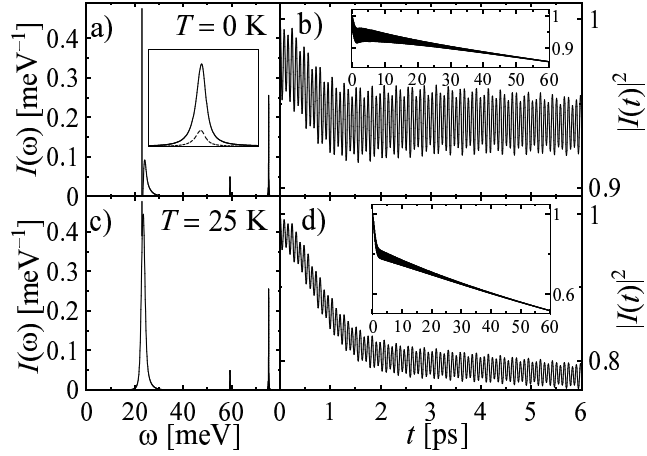}}}
\end{picture}
\end{center}
\caption{\label{fig:sp-dens}
Left: Spectral density vs energy (inset shows the shape of the LO phonon
replica) Right: The corresponding evolution of the
correlation function for two temperatures, as shown (Insets show the
evolution on longer timescales). The
calculation is limited to the two lowest exciton states.}
\end{figure}

Thus, one may write for $I_{0}(\omega)=-(1/\pi){\mathrm{Im}}G_{00}(\omega)$
\begin{eqnarray}
\lefteqn{I_{0}(\omega)  =} \label{sp-dens} \\
&  & Z^{-1}\frac{1}{\pi}\frac{\gamma_{0}/2}%
{(\omega-E_{0})^{2}+\gamma_{0}^{2}/4} \nonumber \\
&  & +\sum_{n,s,\bm{k}}|F^{(\mathrm{s})}_{0n}(\bm{k})|^{2}
\frac{(n_{s,\bm{k}}+1)\delta(\omega-E_{n}-\omega_{s,\bm{k}})
+n_{s,\bm{k}}\delta(\omega-E_{n}+\omega_{s,\bm{k}})}%
{(\omega-E_{0})^{2}+\gamma_{0}^{2}}, \nonumber
\end{eqnarray}
where 
\begin{displaymath}
Z=1-\frac{d\Delta(\omega)}{ d\omega}|_{\omega=E_{0}}
\approx1+\frac{1}{N}\sum_{\bbox{k},s,n'}
\left|\frac{F^{(s)}_{nn'}(\bbox{k})}{E_{n'}-E_{n}+\omega_{s,\bbox{k}}}\right|^{2}
[1+2n_{s,\bbox{k}}].
\end{displaymath}
The Fig. \ref{fig:sp-dens}a,c shows the result.

At $T=0$, $\gamma_{0}(\omega)=0$ at $\omega=E_{0}$ (neglecting
recombination process) and this point is a well defined  pole of the
causal Green function. 
It corresponds to a quasiparticle: the dressed
exciton (i.e. the exciton-polaron, if neglecting LA phonons).
For $T>0$, this spectral density has a Lorentzian
shape around $\omega\sim E_{0}$, with width $\gamma_{0}$. 
Our model is unable to quantitatively account for the broadening
found in experiment. In fact,
to our knowledge, its origin 
has not been explained so far; we will use the experimental
values of $\gamma_{0}$ corresponding to 
630 ps and 170 ps exciton lifetime (including radiative porocesses
and thermally induced transitions) at $T=0$ K and $T=25$ K, 
respectively \cite{borri01}.

Apart from the central peak, the spectral density shows acoustic and
optical phonon sidebands. To the former, only the ground state itself
contributes: the magnitude of the features visible in $\gamma(\omega)$ 
around excited states (Fig. \ref{fig:fnn}c,d) is much smaller than the
energy distance between these states. Therefore, as far as the interaction
with acoustical phonons is concerned, a single level (independent boson)
model supplemented by central line broadening is very accurate.

The features resulting from the LO phonon branch behave in a different
manner: the contribution from nonadiabatic coupling to excited states is
considerable. This results from the fact that the coupling to LO phonons
is much stronger in general, the energies of LO phonons are comparable to
the exciton energy spacing and the coupling to excited levels is stronger
than to the ground state itself due to partial charge cancellation in the
latter. In this case, the contribution from the higher states dominates
over the ground state contribution and the independent boson approach is
not valid.

The time evolution of the dressing process is described by the inverse
Fourier transform of the correlation function given by (\ref{ImGc-cor}).
Taking into account that for $\omega$
in the energy sector of exciton (of order of eV)
$\omega\gg k_{\mathrm{B}}T$, one has
\begin{displaymath}
I_{n}(t)=
- \frac{1}{\pi}\int_{-\infty}^{\infty}d\omega 
{\mathrm{Im}}G_{n,n}(\omega)e^{-i\omega t}.
\end{displaymath}
The time-dependent correlation function may be obtained from (\ref{sp-dens}),
\begin{eqnarray*}
\lefteqn{I_{0}(t)=} \\
& & 
Z^{-1}e^{-i(E_{0}-i\gamma_{0}/2)t} \\
& & 
+\frac{1}{N}\sum_{n,s,\bm{k}}|F^{(s)}_{0n}(\bm{k})|^{2}
\left[ 
\frac{(n_{\bm{k}}+1)e^{-i(E_{n}+\omega_{s,\bm{k}})t}}%
{(\Delta E_{n0}+\omega_{s,\bm{k}})^{2}+\gamma_{0}^{2}}
+ \frac{n_{\bm{k}}e^{-i(E_{n}-\omega_{s,\bm{k}})t}}%
{(\Delta E_{n0}-\omega_{s,\bm{k}})^{2}+\gamma_{0}^{2}}
\right],
\end{eqnarray*}
where $\Delta E_{n0}=E_{n}-E_{0}$.
The result, calculated numerically
for various temperatures, is plotted in 
Fig. \ref{fig:sp-dens}b,d.

The smooth sidebands of $I(\omega)$ 
correspond to the initial correlation decay on 1 ps
timescale. The LO phonon peaks lead to slowly decaying fast oscillations
manifesting phonon beats with frequencies corresponding to the energy
differences $E_{n}-E_{0}-\Omega$. Without including anharmonic
effects, these oscillations would decay very slowly due to the weak LO phonon 
dispersion. When the anharmonic phonon damping (corresponding
to $\tau_{\mathrm{LO}}=9$ ps decay of an LO phonon \cite{vallee91,vallee94})
is included by substituting 
$\omega_{\mathrm{o},\bm{k}}\rightarrow
\omega_{\mathrm{o},\bm{k}}\pm i\gamma_{\mathrm{LO}}$, the LO phonon
beats decay with characteristic time slightly below $\tau_{\mathrm{LO}}$,
which results from the joint effect of damping and dispersion.
The long-time dynamics is governed by the Lorentzian feature
around $\omega=E_{0}$ and shows an exponential decay (as assumed
beforehand).

\section{System response for finite-duration pulses}
\label{sec:trade}

In the previous section we have shown that any change in the carrier
distribution in a QD is followed by lattice relaxation (dressing) processes
which lead to some loss of the quantum coherence. To preserve the
coherence while operating on the carrier states, the action must be
adiabatic with respect to lattice timescales, i.e. phonon periods. Slowing
down the dynamics leads, however, to increasing effect of other
decoherence mechanisms, like radiative recombination or thermally
activated transitions to higher states, as the decoherence caused by such
effects increases linearly with time. 

In order to study the interplay of these decoherence effects we have
studied the evolution of an exciton coupled to acoustic phonons under
optical excitation by a finite-duration laser pulse \cite{alicki03}. 
The effect of
coupling to phonons was quantified in terms of the error of a coherent
operation, $\delta=1-F$, where the fidelity is defined as an overlap
between the actual state and the desired one (the latter was taken to be
the dressed final state, obtained by adiabatic operation), 
$F=\langle\phi|\rho|\phi\rangle$, where $|\phi\rangle$ is the ideal final
state and $\rho$ is the final density matrix. 

The density matrix for the final state is calculated using second order
expansion for the evolution operator of the total system and tracing over
the lattice degrees of freedom (see \cite{alicki03,alicki02a} for details).
For timescales relevant here, it is sufficient to consider acoustic
phonons and only one (ground) exciton state.
The error averaged over initial states can be represented as the overlap 
of two functions
\begin{displaymath}
\delta= \int \frac{d \omega}{\omega^2} R(\omega)S(\omega),
\end{displaymath}
where $R(\omega)$ is the spectral density of the reservoir 
\begin{displaymath}
R(\omega)=\sum_{\mathbf{k}} \left[
\delta (\omega_{\mathbf{k}} -\omega) (n_{\mathbf{k}}+1) 
+ \delta (\omega_{\mathbf{k}} +\omega) n_{\mathbf{k}}	\right]
|F({\mathbf{k}})|^2.
\end{displaymath}
The function $S(\omega)$ represents spectral characteristics 
of the system and is given by
\begin{equation}
S(\omega)=  {1\over 3} (|F_-(\omega)|^2+|F_+(\omega)|^2), \;\;
|F_{\pm}(\omega)|^{2}\approx\alpha^{2} e^{-\frac{1}{2}\tau_{g}^{2}
\left( \omega\pm \frac{\alpha}{\sqrt{2\pi}\tau_{g}} \right)^{2}},
\end{equation}
for a Gaussian driving pulse
$f(t)= {\alpha / (\sqrt{2\pi} \tau_{\mathrm{g}})} e^{-{1\over 2}
\big( {t/\tau_{\mathrm{g}} }\big)^2}$, rotating the state by the angle $\alpha$
on the Bloch sphere.

\begin{figure}[b]
\unitlength 1mm
\begin{center}
\begin{picture}(85,58)
\put(0,0){\resizebox{85mm}{!}{\includegraphics{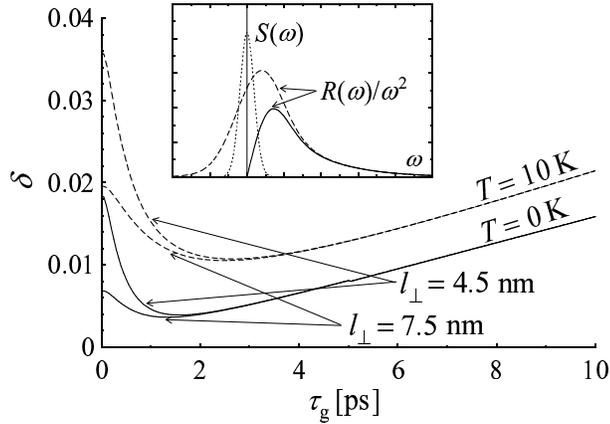}}}
\end{picture}
\end{center}
\caption{\label{fig:result}Combined Markovian and non-Markovian error
for a $\alpha=\pi/2$ rotation on a qubit implemented as a confined
exciton in a InAs/GaAs quantum dot, for $T=0$ (solid lines) and
$T=10$ K (dashed lines), for two dot sizes (dot height is 20\% of its
diameter). The Markovian decoherence
times are inferred from the experimental data \protect\cite{borri01}.
Inset: Spectral density of the phonon
reservoir $R(\omega)$ at these two temperatures and the gate profile
$S(\omega)$ for $\alpha=\pi/2$.}
\end{figure}

For the coupling (\ref{fa}) to LA phonons, one finds at $T=0$ for $\omega\ll
cl_{\perp}^{-1}$
\begin{equation}
R(\omega)\simeq R_{0} \omega^{3},\;\;\;
R_{0}= \frac{(\sigma_{\mathrm{e}}-\sigma_{\mathrm{h}})^{2}}%
{16\pi^{2}\varrho c^{5}},
\end{equation}
and we obtain the averaged error 
$\delta = \alpha^2 R_0 \tau_{\mathrm{g}}^{-2}/3$
(taking into account the upper cut-off, the error will be finite even 
for an infinitely fast gate (see Fig. \ref{fig:result}).

Hence, this non-Markovian error increases as the
speed of gate increases.
This could result in obtaining arbitrarily low error by choosing 
suitably low speed of gates. 
However, if our system is also subjected to other type of noise this
becomes  impossible. Indeed, if we assume that 
our system undergoes additional amplitude damping with 
rate $\gamma_{\mathrm{M}}$, the total error per gate is
\begin{equation}
\delta=\frac{\gamma_{\mathrm{nM}}}{\tau_{\mathrm{g}}^{2}}
+\gamma_{\mathrm{M}}\tau_{\mathrm{g}}, \;\;
\gamma_{\mathrm{nM}}=\frac{1}{3}\alpha^{2}R_{0},\;\;
\gamma_{\mathrm{M}}=\frac{1}{\tau_{\mathrm{r}}},
\end{equation}
where $\tau_{\mathrm{r}}$ is the characteristic time of 
exponential decay (Markovian error).
As a result, the overall error is unavoidable and optimization is needed.
The above formulas lead to 
\begin{displaymath}
\delta_{\mathrm{min}}=
\frac{3}{2}(2\gamma_{\mathrm{M}}^{2}\gamma_{\mathrm{nM}})^{1/3}
=\frac{3}{2}
\left( \frac{2\alpha^{2}R_{0}}{3\tau_{\mathrm{r}}^{2}} \right)^{1/3}
=\alpha^{2/3}0.0035,
\end{displaymath}
for 
\begin{displaymath}
\tau_{\mathrm{g}}=
\left(2\frac{\gamma_{\mathrm{nM}}}{\gamma_{\mathrm{M}}}\right)^{1/3}
=\left( \frac{2}{3}\alpha^{2}R_{0}\tau_{\mathrm{r}} \right)^{1/3}
=\alpha^{2/3}1.5 \mathrm{ps},
\end{displaymath}
where we have used GaAs material parameters (Table \ref{tab:param}).

The exact solution within the proposed model, taking into account the
cut-off and anisotropy of the dot shape and
allowing finite temperatures, is shown in Fig. \ref{fig:result}. The
size-dependent cut-off is reflected by a shift of the optimal
parameters for the two dot sizes: larger dots admit faster gates and
lead to lower error. Interestingly, the trade-off becomes more
apparent at nonzero temperature.

It should be noted that these optimal times are longer than the limits
imposed by level separation 
\cite{biolatti00,chen01,piermarocchi02,tian00,wu02}. 
Thus, the non-Markovian
reservoir effects (dressing) seem to be the essential limitation to 
the gate speed.

\section{Conclusions}

We have shown that an action performed on the carrier system (e.g. exciton)
confined in a quantum dot must be accompanied by an appropriate
reconstruction of the lattice state. Such a dressing process takes place
spontaneously if the charge distribution is changed on times short
compared to timescales of the lattice dynamics (phonon oscillation
periods). This spontaneous process destroys the quantum coherence of the
system and may preclude quantum information processing or coherent
nanotechnology applications. 

On the other hand, sufficiently adiabatic operation would take long time
compared to the exciton lifetime and other exponential decay processes,
again leading to large coherence loss. Therefore, optimal operating
conditions for maximum preservation of coherence must be searched for.

We have evaluated the minimal error for typical parameters, showing
that the optimal gating time ($\sim 1$ ps) lies within the current
experimental possibility. This optimal time
sets up the limit beyond which further gate speed-up is
unfavorable.
It is remarkable that minimal error $\delta$ we found ($0.004-0.01$),
although not extremely high, 
is still of 1-2 orders of magnitude higher than the one  admitted by
fault-tolerant schemes known so far ($\sim 10^{-5}$). However,
possible improvements of the latter schemes cannot be excluded.

\begin{acknowledgments}
This work was supported by the Polish KBN under Grants No. 2 P03B 024 24
and No. 2 P03B 085 25.
\end{acknowledgments}

\bibliographystyle{unsrt}
\bibliography{abbr,quantum}

\end{document}